%Paper: astro-ph/9412035
%From: hogan@astro.washington.edu (Craig Hogan)
%Date: Fri, 9 Dec 94 09:53:52 -0800
%Date (revised): Fri, 9 Dec 94 11:20:29 -0800

\headline={\ifnum\pageno=1\firstheadline\else
\ifodd\pageno\rightheadline \else\leftheadline\fi\fi}
\def\firstheadline{\hfil}
\def\rightheadline{\hfil}
\def\leftheadline{\hfil}
        \footline={\ifnum\pageno=1\firstfootline\else\otherfootline\fi}
\def\firstfootline{\rm\hss\folio\hss}
\def\otherfootline{\hfil}

\font\tenrm=cmr10
\font\tenit=cmti10
\font\elevenbf=cmbx10 scaled\magstep 1
 1
 1

%\TagsOnRight
\nopagenumbers
\line{\hfil }
\vglue 1cm
\hsize=6.0truein
\vsize=8.5truein
\parindent=3pc
\baselineskip=10pt
\newcount\fcount \fcount=0
\def\ref#1{\global\advance\fcount by 1
\global\xdef#1{\relax\the\fcount}}
\def\refs{\hangindent=5ex\hangafter=1}

\def\msol{{\,\rm M_{\odot}}}
\def\lsol{{\,\rm L_{\odot}}}

\centerline{\elevenbf
Primordial Deuterium Abundance and   Cosmic Baryon Density}
\vglue 1.0cm
\centerline{\tenrm CRAIG J. HOGAN }
\baselineskip=13pt
\centerline{\tenit Astronomy and Physics Departments, University of Washington}
\baselineskip=12pt
\centerline{\tenit Seattle, WA 98195, USA}
\vglue 0.3cm
\centerline{\tenrm ABSTRACT}
\vglue 0.3cm
{\rightskip=3pc
 \leftskip=3pc
 \tenrm\baselineskip=12pt%\parindent=1pc
 \noindent
The comparison of cosmic abundances of the light elements with the density of
baryonic stars and gas in the universe today provides a critical test of
big bang
theory  and  a powerful probe of the nature of  dark matter.    A new
technique allows
determination  of cosmic deuterium abundances in
quasar absorption clouds
at large redshift, allowing a new test of big bang homogeneity in diverse,
very distant
systems.  The first results of these studies
are summarized, along with their implications. The quasar data are
confronted with the
apparently contradictory story from the helium-3 abundances measured in our
Galaxy.
The density of baryonic stars
and gas in the universe today  is reviewed and compared with the big bang
prediction.
\vglue 0.8cm }
\line{\elevenbf 1. Primordial Deuterium and Galactic $^3$He \hfil}
\bigskip
\baselineskip=13pt
\tenrm
People never seem to tire of telling or hearing the triumphant story of the
cosmic
light element abundances--- for example,  about how the observed predominance
of
hydrogen, and the  helium mass fraction of roughly 25\%, confirm the basic
tenets of the
big bang theory back to a time of one second, or about how the precisely
determined
$^4He$ mass fraction in low-metal galaxies, $0.228\pm 0.005$, constrains
the cosmic
baryon density to be less than $\Omega_b=0.015 h^{-2}$. Here however I will
disappoint the avid listeners by focussing attention on
the more ambiguous story of the primordial deuterium abundance.

The cosmic baryon-to-photon  ratio $\eta$ is the one astrophysical
parameter of big
bang nucleosynthesis theory. Everything else follows from pure physics,
including
predictions for the abundances of light elements.
Since we know the photon density today, and since the entropy has changed
rather
little since nucleosynthesis (as indicated by the perfect Planck spectrum of
the
background radiation),
$\eta$ determines the physical density of baryons today. In units of the cosmic
critical density, the density of baryons is
$$
\Omega_b=3.73\times 10^{-3}h^{-2}\eta_{10}
$$
where $\eta_{10}=10^{10}\eta$. The theory
predicts\ref\terry\ref\smith[\terry,\smith]
a mass fraction of $^4He$
$$
Y_p=0.228+0.01\eta_{10}
$$
and a primordial number ratio of deuterium to hydrogen
$$
(D/H)_p=4.6\times 10^{-4}\eta_{10}^{-5/3}
=2.5\times 10^{-4}\left({\Omega_bh^2\over 0.005}\right)^{-5/3}
=4.5\times 10^{-5}\left({\Omega_bh^2\over 0.015}\right)^{-5/3},
$$
where the last two ``high'' and ``low'' values are chosen to illuminate the
arguments
below. The helium is the best verification of the big bang theory, but the
deuterium is
the best measure of baryon density, since it is fairly sensitive to $\eta$.

The trick is to measure the primordial abundance of this fragile element.
Since deuterium is almost impossible to make outside of the big bang,
almost all cosmic deuterium is a relic of primordial
nucleosynthesis, so it is easy to get a lower limit on $(D/H)_p$. But since
the big
bang, successive generations of stars have destroyed most  of it by burning
to heavier
elements, which makes any upper limit less straightforward. Indirect
constraints on
$(D/H)_p$  have been deduced from $D$ abundances in the reprocessed
material in the
Galactic interstellar medium and from
 the local abundances of the principal immediate product
of $D$ burning, $^3He$.
Using models of galactic chemical evolution (a weak spot in the argument),
the solar
abundance of
 $^3He$ is used to give an upper limit
on the primordial $(D/H)_p$,  and thereby  a lower limit on
the  $\eta$ from standard big bang
nucleosynthesis theory[\terry,\smith].
This  lower limit is  higher  than observed density of baryons, and
therefore  provides an argument for abundant baryonic dark matter.  If this
lower
limit is relaxed, the range of allowed
 $\eta$ includes lower values where the predicted $^4He$ abundance lies
more comfortably close to observations,  where the bound on the
number of particle species is  relaxed, and where there is almost no
baryonic dark matter.

Advances in instrumentation and telescope aperture, especially the Keck
telescope and the HIRES
spectrograph, have recently enabled observations of high-redshift quasars
at high spectral
resolution, of the order of
$10 {\rm km\ sec^{-1}}$. This has enabled detailed study of the Lyman series
absorption lines of  hydrogen and its deuterium counterpart (shifted by one
part in
3700 to the blue) in high-column-density foreground
absorbers\ref\toni\ref\cars[\toni,\cars].  The first such observations
revealed
absorption consistent with a high abundance of deuterium in  one very
metal-poor absorbing cloud in
one quasar (Q0014+813), around $D/H\approx 2\times 10^{-4}$ by number,
which is about a factor
of five higher than previous guesses for the primordial
$D/H$ based on observations of local $^3He$ abundances, and an order of
magnitude
greater than the interstellar value\ref\linsky[\linsky]. As of this writing,
the
abundance in one other, slightly more enriched cloud has been
measured\ref\tytler[\tytler]   closer
to the ISM value.
One Ly series absorption line system can always be dismissed
as a chance hydrogen interloper masquerading as
deuterium\ref\steigman[\steigman], while a low D/H
can be attributed to  destruction, so    we require more systems for a
clearer picture to
emerge of the true primordial value. The technique will in time  be applied
to   many other
quasars and absorbers, establishing critical tests we have never had before
of big bang
theory: abundances at cosmological distances, in different environments and at
different cosmic epochs, and in  pristine material which has undergone
relatively
little chemical evolution (see figure 2). If more than one high $D/H$ value
is found,
it will lead to a major revision of thinking about this isotope.
While not conclusive, the first data have raised at least the possibility
that the primordial
$D/H$ might be rather high, and the global baryon density rather low,
motivating a
reexamination of the interpretation of cosmic abundances and the data on
the density of
baryons.

The  measure of $^3He/H$ most widely used  for estimating $(D/H)_p$
comes from
measurements of the solar wind, both from direct exposure experiments
and from meteorites, of   $^3He/ ^4He$\ref\geiss[\geiss], which
  are used to infer that  the   abundances of
the presolar nebula
  by number
were $D/H=2.6\pm 1.0\times 10^{-5}$,
$^3He/H=1.5\pm 0.3\times 10^{-5}$,
$(D+\ ^3He)/H=4.1\pm 1.0\times 10^{-5}$.
These are  taken as Galactic or solar-circle
averages for the purpose of defining constraints on primordial abundances.
A skein of theory then connects the presolar abundance with the primordial
one.
The main reason why the new Keck $D/H$ observations come as a surprise was that
they give a $D/H$ a factor of five larger than
$^3He$ abundances in the solar system.

\vfil\eject
\null
\vskip 6.5in
%\special{picture lightcone}
\vglue 0.3cm

{{
\rightskip=1pc
 \leftskip=7pc
 \tenrm\baselineskip=10pt%\parindent=1pc
\noindent
Figure 1-- Comoving lightcone in an
Einstein-deSitter model, showing spacetime location
of Q0014+83 and the ``Chaffee cloud'' where
Songaila et al. estimated $D/H$ from Lyman series
absorption. Vertical position corresponds to cosmic scale factor;
the bottom of the cone is $t=0$,
the top is the present epoch.  Horizontal position is the present-day proper
distance; the apex is our position, and
the classical particle horizon is the
bottom of the cone.  The great distance of the cloud
tests the cosmological principle  applied to
primordial abundances,  the first  measurement probing the early  past
world lines of distant points--- the distant interior of our past light cone.
The high redshift of the absorber provides a direct probe of
pristine material from the big bang; this cloud has metal abundances
estimated at less than
$10^{-3}$ of the solar value.
\hfil\vfil }}
\vfil\eject

In the standard picture    the bulk
of   primordial $D$  in the Galaxy is
burned to $^3He$ in protostellar  collapse. Galactic chemical evolution
models\ref\gary\ref\vang\ref\galli[\gary,\vang,\galli]
 show that $D/H$ can
be reduced in this way to its present interstellar value ($\approx
1.5\times 10^{-5}$,
ref. \linsky) from any plausible initial
value. However, in the low mass
stars which now dominate the chemical recycling of the interstellar medium
(ISM),
the bulk of the material is assumed to be never heated to the higher
temperature
required to burn the $^3He$, so the bulk of
the primordial $D$  reappears
in the ISM as $^3He$ when the envelopes are ejected.
 For the galaxy as a whole, the sum
$(D+ ^3He)/H$ therefore only increases with time,  so that
even the solar $^3He$ abundance can be used to set constraints on $(D/H)_p$.
 This is why the evolution of $^3He$ is critical.\ref\helium[\helium]

The only other useful  measure of cosmic $^3He/H$
 comes from  radio
emission maps  of highly ionized HII regions in the
Galaxy\ref\balser\ref\wilson[\balser,\wilson].  The column density of
$^3He^+$ is
estimated from the brightness in the
 8.665 GHz hyperfine transition line, and the column
(squared) density of
$H^+$ or $^4He^+$ is estimated from   radio recombination lines.
Balser et al. use this data and a simple  model of the gas distribution
to obtain reliable estimates of $^3He/H$ in 7 Galactic HII regions,
and ``preliminary'' abundances and limits in 7 more.
Two of the most reliable ones are W43 and W49, with low values
$^3He/H= 1.13\pm 0.1 \times 10^{-5}$ and
 $^3He/H= 0.68\pm 0.15\times 10^{-5}$ respectively.
There appears to be a real  range of values, with W3 for
example measured at  $^3He/H= 4.22\pm 0.08\times 10^{-5}$,
and some are consistent with still higher values.
There may be a trend with galactocentric radius in the sense that
lower values tend to lie within the solar orbit and higher
values outside it.

Note that these results do not mesh with the standard interpretation of the
solar
$^3He$;
quite aside from the Keck D/H  observations,   empirical evidence
in the Galaxy suggest  that
stellar populations on average actually get rid of $^3He$.

\item{$\bullet$} The
Solar System value $(D+\ ^3He)/H$ is   greater than the interstellar one; if
$(D+\ ^3He)/H$ were  steadily increasing, it ought to be less, because of
the elapsed time since the formation of the solar system.

\item{$\bullet$} The
ISM shows large variations in $^3He/H$, which argues that one ought
not take any one point, such as the solar system, as an average
of the Galactic abundance, and that simple uniform-mixing models are
unlikely to
accurately model the abundances at any given point, such as the solar system.

\item{$\bullet$} The
gradient with Galactic radius goes the wrong way; if stars are creating
$^3He$ on average, it ought to be highly enriched towards the Galactic
center, like other heavier elements are.

\item{$\bullet$} If we adopt instead the lowest $^3He/H$ value in ISM (W49)
as the
primordial one, to be consistent with the idea that $^3He/H$ cannot
decrease, thereby
assuming that the additional   $^3He$ found at other sites is Galactic in
origin
as required in the standard picture, {\it then}
 SSBN requires a large $\Omega_bh^2=0.075$,
in which case it also predicts an excessive $^4He$ abundance
$Y_p\approx 0.26$. The   observed value is $Y_p=0.228\pm 0.005$
\ref\pagel\ref\skill[\pagel,\skill], which is marginally
inconsistent even with the SBBN prediction for solar $^3He/H$, $Y_p=0.242$.

  A destruction mechanism is therefore
desirable  both  for improving the consistency of big bang theory and for
interpreting the Galactic $^3He/H$ data.
It is not clear whether such a mechanism operates in the Galaxy.
One recently proposed
  mechanism[\helium] is based on mixing envelope material in low mass stars
down to high temperature after they reach the giant branch, so that the
$^3He$ is
destroyed before the material is ejected. This process,
 originally postulated to explain the  observed change in  C and N isotope
abundances\ref\char[\char,\ref\brown \brown] as low mass stars ascend the
giant branch,
 would also destroy  $^3He$. It remains to be seen how  important it is for
the population as a
whole, but the possibility of such effects motivates caution in using
highly processed material
for estimating $(D/H)_p$.

\bigskip
\line{\elevenbf 2. Cosmic  Baryon Bookkeeping\hfil}
\bigskip

It is interesting to compare the density of baryons inferred from either SBBN
argument with the
density of baryons and dark matter found in the universe.

Let us review a number of different measures
of global densities, summarized in figure 2. Each column
shows both estimated statistical errors in
the method and the  variation
with the (still uncertain) Hubble constant $h$,
where $h=H_0/100 {\rm km\ sec^{-1}\  Mpc^{-1}}$.

The first column shows the  contribution of
baryons to the mean density,  estimated from standard Big Bang
nucleosynthesis (SBBN).  The current canonical
 range[\terry,\smith,  \gary] is shown,
$\Omega_bh^2=0.010-0.015$,  which leads to
the best concordance with a low value of $(D/H)_p$. Most reliable
 is the
upper limit of this range, which appears firmly fixed by
a variety of abundances.  Indeed, it represents
 a $3\sigma$ departure from the best value[\pagel]
for the primordial $^4He$ abundance, $Y_P=0.228\pm 0.005.$
Some have argued\ref\jean[\jean] that one
should instead fit the best value of $Y_P$, requiring
 $\Omega_bh^2\approx 0.005$,  the lower indicated range.
 This    estimate
 of $\Omega_bh^2$  agrees with the recent
possible detection[\toni] of   high deuterium.

 A direct lower limit on gas density is imposed by
quasar absorption line statistics, shown in the second
column. A large sample of quasars
provides  an accurate census of all neutral hydrogen
in the universe at $z<4$ through   Ly$\alpha$ absorption along their
sightlines.
  At high redshift, the bulk of
the HI is in high column density damped Lyman $\alpha$
(DLy$\alpha$) absorbers, with column density in the range
$N(HI)\approx 10^{20-22}{\rm cm^{-2}}$.
These contribute\ref\wolfe[\wolfe] an integrated
 density $\Omega_{\rm DLy\alpha}h
= 2.9\pm 0.6\times 10^{-3}$, which should be taken as a lower limit
on the total density of such absorbers\ref\fall[\fall].

The third column shows an example of traditional
baryonic bookkeeping\ref\bt[\bt]: estimate the mass-to-light
ratio of a population, then use the mean cosmic luminosity
density (here, in $V$ band and solar units) to find a contribution to the
mean mass
density.  We use spiral galaxies, as they dominate the luminosity
density. Our Galactic disk  out to 700pc height has $(M/L)_V=5\msol/\lsol$,
with most of the mass contained in stars.
 If  all spiral  disks have the same mass-to-light ratio,
we can use an estimate[\bt] of the luminosity density
($j_0=1.7\pm 0.6\times 10^8\lsol/{Mpc^3}$ in $ V$) to
get the integrated density of all
the material in spiral galaxy disks.
(Note that the errors shown are just those from the estimate of $j_0$).
Similarly there are several mass measurements of
the Galaxy halo mass from local group
 satellite galaxy orbits\ref\zaritsky[\zaritsky]
and from local group timing,
which yield masses of at least
$1 \times 10^{12}\msol$. With
a  Galactic luminosity[\bt] of
$1.4\times 10^{10}\lsol$ this  corresponds to an overall
lower limit of
$(M/L)_V=71\msol/\lsol$ for the Galaxy.
 If we assume that
  all spirals have a similar amount of halo
material per  disk light, we
obtain the estimate shown for spiral halos,
$\Omega_{\rm halos}h> 4.4\pm 1.5\times 10^{-2}$.

A variant of this argument is shown in column four. Persic and
Salucci\ref\persic[\persic]  have integrated the luminosity functions of
spiral and
elliptical galaxies separately, with  their $M/L$ estimated directly from
dynamics,
allowing $h$ to be eliminated. They estimate $\Omega_b=1.5\times 10^{-3}$ from
ellipticals and
$\Omega_b=0.7\times 10^{-3}$ from spirals, which is systematically lower, and
probably more accurate, than the previous argument. I have estimated
errors to again
be at the 30\% level. It is interesting that the total contribution from gas in
groups and clusters is comparable to these, $\Omega_b=1.5\times
10^{-3}h_{50}^{-1.3}$, in spite of the fact that rich clusters contain only
about
1\% of galaxies, and that the clusters out to the Abell radius were
assembled from about 1\% of the total comoving volume, but there ought to be
large errors attached to this
estimate.

\vfil\eject
\null
\vskip  4.2in
%{\special{picture baryons}}
\null
\vglue 0.3cm
{\rightskip=0pc
 \leftskip=8pc
 \tenrm\baselineskip=11pt%\parindent=1pc
 \noindent
Figure 2-- Estimated contributions of various
components to the global density, in units of the
cosmic critical density. Each column shows a
vertical range from estimated internal errors, as
well as a variation across each column   due to the
range of possible values of the Hubble constant,
$0.5<h<1$. Column 1  shows the  range for $\Omega_b$
allowed by SBBN, both for the canonical limits
derived from solar system $^3He$ abundance, and for
 the lower value estimated from
the recent possible detection of high primordial $D/H$ in a single QSO absorber
and the best measured value of
primordial $^4He$ abundance
$Y_p$.
Column 2 shows the contribution of neutral hydrogen in quasar
${\rm DLy\alpha}$ absorbers at $ z\approx 3.5$; although Wolfe's errors are
shown,
this should be taken as a lower limit for the HI density. Column 3
shows  estimates of the global density of spiral galaxy disk stars
and halos, obtained from local (Galactic) $M/L$ estimates combined with
the mean cosmic luminosity density. The lower band represents
$M/L=5\msol/\lsol$, representative of the local disk material
out to about 700 pc,
and the upper band ($\Omega_{\rm halos}$)
 represents $M/L=71\msol/\lsol$, corresponding
to a Galaxy halo mass of $1\times 10^{12}$ solar masses.
 Column  4  shows a similar estimate but based on integration of luminosity
functions
 and
dynamically estimated $M/L$; the band represents Persic and Salucci's
estimate for
spiral and elliptical  galaxies,
with errors added by me, and the upper line represents their estimate
including cluster
gas. Columns 5
and 6 show    global densities estimated
from the ratio of components in the Coma cluster. Column 5
shows the  dark matter, gas and star components where the sum
is assumed to have $\Omega=1$; Column 6 shows the same ratios,
but where the total density is fixed to be 0.1.
 \hfil\vfil}
\bigskip
\tenrm

\baselineskip=13pt
\tenrm

Column five is based on just the Coma cluster, where we have the most
uniformly reliable
data\ref\white[\white]: it shows cosmic densities based on
the estimated mass for Coma
($M=1.1\pm0.18\times 10^{15}h^{-1}\msol,
M_{gas}=5.45\pm 0.98\times 10^{13}h^{-5/2}\msol,
M_{stars}=1.0\pm 0.2\times 10^{13}h^{-1}\msol$),
 assuming that  Coma is representative of their cosmic ratios
and that
$\Omega_{PDM}=1$.
 The high baryon density is in  apparent
conflict with SBBN.
Ironically,
the high $M/L$ of galaxy clusters,  regarded since Zwicky  as   the strongest
evidence for plentiful cosmic dark matter, is now apparently
due to low $L$ (i.e., the bulk  of the baryons being in gas rather
than stars),  and not to high $M$ (i.e., having a more
representative sample of the high cosmic dark matter
density.) The well-studied central
region of Coma   has about the same mix of visible
baryons and dynamical dark matter as the Milky Way. This is shown in
column  five, which  again shows  the empirical
 ratios of mass to gas and stars in the Coma cluster, only now
assuming instead that $\Omega=0.1$ for the dark matter.

Where then are the baryons?
Common thinking   is that most of the
baryons reside in a photoionized IGM--- everywhere outside of clusters,
there is the
same large amount of gas per galaxy as there is inside, but it is not seen
because it is
not hot and dense enough to emit X-rays.  But it
remains difficult to reconcile the large number of cluster baryons
 with the small
$\Omega_b$ required by   even the highest SBBN limits.
Galaxy halos
contain about the same amount of material as
 the   total baryonic density  for   canonical (low D/H) SBBN;
a popular option is to make the halos out of compact objects (MACHOs; see
Carr's contribution
in this volume).
But making halos out of MACHOs prevents us from using these baryons in
the IGM;   baryons would have to be converted into MACHOs
in galaxies, and gas in clusters, so the formation of the MACHOs would need
to be at a low redshift,
accessible to observation.
The  lower value
of $\Omega_bh^2=0.005$
does  not   provide even enough baryons to make galaxy halos.
There is no easy way to reconcile the large baryon abundance of clusters
with this very low baryon density; a cluster like Coma would need  to gather
  baryons from a volume twenty times bigger than the volume from which it
gets galaxies.

One route to reconciliation is the introduction of inhomogeneities in the
baryon
distribution within the early big bang,  on either small or large scales.
Small scale inhomogeneities, which can arise naturally from QCD or
electroweak phase
transitions,  create zones of  neutron-rich
nucleosynthesis\ref\apple\ref\texas\ref\malaney\ref\fuller [\apple, \texas,
\malaney,
\fuller], which naturally creates extra deuterium[\apple]
compared to a homogeneous model; these models however do not appear capable
of exceeding
the usual SBBN upper limit[\fuller].
There could also be more baryons if some of them gravitationally collapse into
a
compact form at or before recombination, due to large-scale inhomogeneities
 \ref\hogan
\ref\gor[\hogan,\gor]. Some noise distributions   naturally produce
approximately correct
abundance distributions in the diffuse baryons that do not collapse,
so in principle the total density of baryonic matter   could be   greater than
classical nucleosynthesis bounds, although they cannot easily solve the
Coma cluster problem,
which necessarily involves diffuse gas.  Such ideas can be constrained or
eliminated by a variety of
gravitational microlens probes of the nature of dark
matter\ref\macho\ref\eros\ref\dal\ref\haw\ref\crotts[\macho,\eros,\dal,\haw,
\crotts],
and  can also be tested directly, by measuring abundances in different
locations  (eg, different
quasar absorbers.)
It is important to emphasize that the overall concordance between the
different light element
abundances remains good, which indicates that the basic SBBN picture is a good
approximation--- and that there is no reason to think that this situation
will change.

This work was supported by NASA grant  NAGW-2523 and NSF grant AST 9320045
at the University of
Washington.

 \vfil
 \eject
\line{\elevenbf 5. References \hfil}
\vglue 0.4cm
\def\bib{\noindent\hang}
\def\apj{ApJ}

\bib\refs\terry .   Walker, T. P., Steigman, G., Schramm, D. N.,
Olive, K. A., Kang, H. S. {\it Astrophys. J.}, {\bf 376}, 51 (1991).

\bib\refs\smith . Smith, M. S., Kawano, L. H. \& Malaney, R. A.
{\it Astrophys. J. Suppl.}, {\bf 85}, 219 (1993).

\bib\refs\toni . Songaila, A., Cowie, L. L., Hogan, C. J., \& Rugers,
M. {\it Nature}, submitted (1994)

\bib\refs\cars . Carswell, R. F., Rauch, M., Weymann, R. J.,  Cooke, A. J.,
and Webb, J. K., 1994, MNRAS, 268, L1-L4

\bib\refs\linsky . Linsky, J. L., et al., 1992, \apj, 402, 694

\bib\refs\tytler . Tytler, D., and Fan, X., private communication

\bib\refs\steigman . Steigman, G.,  1994, MNRAS, 269, L53

\bib\refs\geiss .  Geiss, J. 1993, in Origin and Evolution of the Elements,
ed. N. Prantzos,
E. Vangioni-Flam and M. Cass\'e (Cambridge: Cambridge Univ. Press), p. 89

\bib\refs\gary . Steigman, G. \& Tosi, M. {\it Astrophys. J.}, {\bf 401},
150 (1992).

\bib\refs\vang . Vangioni-Flam, E., Olive, K. A., and Prantzos, N., 1994,
\apj, 427,618

\bib\refs\galli .  Galli, D., Palla, F.,   Ferrini, F., Penco, U., 1994,
ApJ in press, Arcetri preprint 20/94 %big evolution paper

\bib\refs\helium . Hogan, C. J., 1994, {\it Astrophys. J.}, submitted

\bib\refs\balser . Balser, D. S., Bania, T. M., Brockway, C. J., Rood, R.
T., and Wilson,
 T. L., 1994, \apj, in press

\bib\refs\wilson . Wilson, T. L., and Rood, R. T., 1994, Ann. Rev. A. Ap.,
32, in press

\bib\refs\pagel . Pagel, B. E. J., Simonson, E. A., Terlevich, R. J., \&
Edmunds, M. G., {\it Mon. Not. Roy. astr. Soc.},  {\it 255}, 325-345 (1992)

\bib\refs\skill . Skillman, E. D., and Kennicutt, R. C., 1993, \apj, 411, 655

\bib\refs\char .  Charbonnel, C. 1994, AA, 282,811

\bib\refs\brown . Brown, J. A., Wallerstein, G. W., and Oke, J.B. 1990, AJ
, 100, 1561

\bib\refs\jean . Vangioni-Flam, E. \& Audouze, J. {\it Astron.
Astrophys.}, {\bf 193}, 81 (1988).

\bib\refs\wolfe . Wolfe, A. M.
{\it Ann. N. Y. Acad. Sci.}, {\bf 688}, 281 - 296 (1993).

\bib\refs\fall . Fall, S. M. \& Pei, Y. C.,
 {\it Astrophys. J.}, {\bf 402}, 479-492 (1993)

\bib\refs\bt . Binney, J. \& Tremaine, S., {\it Galactic Dynamics},
Princeton University Press, (1987)

\bib\refs\zaritsky .  Zaritsky, D.,  Olszewski, E. W.,  Schommer, R. A.,
Peterson, R. C.,
              and  Aaronson, M. {\it Astrophys. J.},
 {\bf 345}, 759-769 (1989)

\bib\refs\persic .  Persic, M. and Salucci, P., 1992, MNRAS, 258, 14P

\bib\refs\white . White, S. D. M., Navarro, J. F.,
Evrard, A. E., \& Frenk, C. S., {\it Nature},{\bf 366}, 429-433 (1993)

\bib\refs\apple . Applegate, J. H., and Hogan, C. J., 1985, Phys. Rev. D
30, 3037

\bib\refs\texas . Hogan, C. J., 1991, Ann. N.Y. Acad. Sci., 1991, 647, 76

\bib\refs\malaney . Malaney, R. A., and Mathews, G. J., 1993, Phys Rep 229, 145

\bib\refs\fuller . Jedamzik, K.,  Fuller, G., and Mathews, G., 1994, ApJ,
in press

\bib\refs\hogan . Hogan, C. J. {\it Astrophys. J.}, {\bf 415}, L63 - L66 (1993)

\bib\refs\gor . Gnedin, N. Y., Ostriker, J. P., and Rees, M. J., 1994, ApJ,
in press

\bib\refs\macho . Alcock, C., et al.,
 \ {\it Nature}, {\bf 365}, 621-623 (1993).

\bib\refs\eros . Aubourg, E., et al.,
 \ {\it Nature}, {\bf 365}, 623-625 (1993)

\bib\refs\dal . Dalcanton, J. J., Canizares, C. R.,
Granados, A., Steidel, C. C., and Stocke, J. T., {\it Astrophys. J.},
424, 550 (1994).

\bib\refs\haw . Hawkins, M. R. S., {\it Nature}, 366, 242 (1993).

\bib\refs\crotts . Crotts, A. P., {\it Astrophys. J.},
{\bf 399}, L43-L46, (1992)

\vfil
\eject
\bye